\documentclass[11pt]{article}
\usepackage{algorithm}
\usepackage{algpseudocode}
\PassOptionsToPackage{hyphens}{url}
\usepackage{hyperref}

\usepackage{amssymb}
\usepackage{amsfonts}
\usepackage{amsmath}
\providecommand{\U}[1]{\protect\rule{.1in}{.1in}}
\usepackage{tikz}
\usetikzlibrary{shapes}
\usepackage{pgfplots}
\usepackage{verbatim}
\usepgfplotslibrary{statistics}
\pgfplotsset{compat=1.8}

\usepackage[caption=false]{subfig}
\usepackage{graphicx}

\usepackage[normalem]{ulem}

\renewcommand\and{\end{tabular}\kern-\tabcolsep\ and\ \kern-\tabcolsep\begin{tabular}[t]{c}}
\let\origthanks\thanks
\renewcommand\thanks[1]{\begingroup\let\rlap\relax\origthanks{#1}\endgroup}

\newcommand{\quotes}[1]{``#1''}

\begin{document}

\title{Mobile smartphone tracing can detect almost all SARS-CoV-2 infections}
\author{Bastian Prasse\thanks{Faculty of Electrical Engineering, Mathematics and
Computer Science, P.O Box 5031, 2600 GA Delft, The Netherlands; \emph{email}:
b.prasse@tudelft.nl, p.f.a.vanmieghem@tudelft.nl} \and Piet Van Mieghem\footnotemark[1]}
\date{Delft University of Technology\\
 September 21, 2020}
\maketitle

\begin{abstract}
Currently, many countries are considering the introduction of tracing software on mobile smartphones with the main purpose to inform and alarm the mobile app user. Here, we demonstrate that, in addition to alarming and informing, mobile tracing \textit{can detect nearly all users that are infected by SARS-CoV-2}. Our algorithm BETIS (Bayesian Estimation for Tracing Infection States) makes use of self-reports of the user's health status. Then, BETIS guarantees that almost all SARS-CoV-2 infections of the group of users can be detected. Furthermore, BETIS estimates the virus prevalence in the \textit{whole} population, consisting of users and non-users. BETIS is based on a hidden Markov epidemic model and recursive Bayesian filtering. The potential that mobile tracing apps, in addition to medical testing and quarantining, can eradicate COVID-19 may persuade citizens to trade-off privacy against public health.
\end{abstract}

\section{Introduction}
The COVID-19 pandemic triggered firm lockdowns of societies and economies around the world. Lockdown measures must be released gently and, if necessary, retightened to avoid a dramatic second wave of COVID-19. To trace the pandemic, smartphone apps have recently received a lot of attention \cite{ferretti2020quantifying, oliver2020mobile, drew2020rapid}. A particular challenge to estimating the prevalence of COVID-19 are the asymptomatic infections. Recent \textit{contact apps} aim to alarm the user of a potential infection, if the user has been close to another user with a confirmed SARS-CoV-2 infection. Alarming individuals by contact apps is a particular method of \textit{social alertness} \cite{funk2010modelling, kiss2010impact, sahneh2011epidemic, sahneh2012existence, theodorakopoulos2012selfish}. If alerted, individuals are more cautious and less likely to become infected. For a comparison of the effect of social alertness and social distancing, we refer the reader to \cite{schumm2013impact}.The awareness of potential infections may lead to suppression of the virus~\cite{funk2009spread}.

The intended use of some smartphone app goes beyond alarming individuals. For instance, in the \textit{COVID Symptom Study} \cite{drew2020rapid}, smartphone users provide their health status as a self-report via an app on a daily basis. The self reports include user information, such as age and location, and potential COVID-19 symptoms, such as fever or loss of smell and taste. The self-reports aid at identifying emerging geographical hotspots of SARS-CoV-2 infections.

Previous studies \cite{tizzoni2014use, bengtsson2015using, finger2016mobile, oliver2020mobile} consider \textit{aggregated} location information, in the form of mobility flow or population density. Here, we explore the full potential of location information for tracing the spread of COVID-19. More precisely, given the locations of the app users, our algorithm called BETIS, Bayesian Estimation for Tracing Infection States, finds nearly all infected users. Furthermore, BETIS traces the total number of infections in the \textit{whole} population, consisting of users and non-users. Hence, complementing BETIS with boarder control, medical testing and quarantine enforcement is a second potential pillar, besides vaccine development, to eradicate the coronavirus. Since society seems convinced that the only hope to abandon the destructive impact of COVID-19 is a vaccine, we believe that BETIS is a worthy second horse in the race.

\section{Epidemic model}\label{sec:model}

We consider the spread of SARS-CoV-2 among $N$ individuals. The individuals $i = 1, ..., N_\text{u}$, with $N_\text{u} \le N$, are users of the smartphone app. Thus, the fraction of smartphone users equals $c_0 = N_\text{u}/N$, while the remaining individuals $N_\text{u} + 1, ..., N$ do not use the app. Every user $i=1, ..., N_\text{u}$ reports COVID-19 related symptoms through the app, e.g., via a questionnaire \cite{drew2020rapid}. At any discrete time $k\in \mathbb{N}$, every individual $i$ has a viral state $X_i[k]\in \mathcal{C}$. The set of compartments equals $\mathcal{C}= \{ \mathcal{S},\mathcal{S}_\text{fa},\mathcal{E}, \mathcal{I},\mathcal{I}_\text{a}, \mathcal{R}\}$. The state $X_i[k]=\mathcal{S}$ denotes that individual $i$ is \textit{susceptible} (healthy). There are other diseases with similar symptoms as COVID-19, for instance influenza. Thus, the self-reports via the app might produce \textit{false alarms}, which point erroneously to a SARS-CoV-2 infection while the individual suffers from another disease. The viral state $X_i[k]=\mathcal{S}_\text{fa}$ indicates that individual $i$ is infected by a disease other than COVID-19 with similar symptoms. The \textit{exposed} state $X_i[k]=\mathcal{E}$ denotes that individual $i$ is infected by SARS-CoV-2 but not contagious yet. After the exposed state $\mathcal{E}$, an individual becomes either \textit{infectious symptomatic} $\mathcal{I}$ or \textit{infectious asymptomatic} $\mathcal{I}_\text{a}$. Individuals in either infectious state $\mathcal{I}$ and $\mathcal{I}_\text{a}$ are contagious to susceptible individuals in their vicinity. After some time, symptomatic infected individuals in $\mathcal{I}$ transition to the \textit{symptomatic removed} state $\mathcal{R}$, due to recovery, quarantine, hospitalisation or death. Removed individuals in $\mathcal{R}$ cannot infect susceptible individuals any longer. We assume that a recovered individual is immune. Hence, multiple infections do not occur. 

 The BETIS algorithm estimates the viral states $X_i[k]$ of an app user $i$. Additionally to the health self-reports, BETIS uses the neighbourhood $N_{\text{u},i}[k] \subset \{1,...,N_\text{u}\}$ for each user $i = 1,...,N_\text{u}$. The neighbourhood $N_{\text{u},i}[k]$ consists of the contacts of user $i$ to other users at time $k$. Two users are \quotes{in contact} with each other, if the users are physically close for a sufficiently long time period. For instance, the NHS Test and Trace service define a contact when users are within 2 meters of each other for more than 15 minutes \cite{UK_NHS_test_and_trace}. The neighbourhood $N_{\text{u},i}[k]$ can be obtained in two ways: The mobile app can perform direct measurements of the neighbourhood $N_{\text{u},i}[k]$, e.g., by Bluetooth. Alternatively, the app can use a $2\times 1$ location vector $z_i[k]\in \mathbb{R}^2$, which specifies the latitude and longitude of user $i$ at time $k$ and can be obtained, for instance, by GPS. The neighbourhood of user $i$ is obtained from the location vector $z_i[k]$ by 
\begin{align*}
\mathcal{N}_{\text{u}, i}[k] =\left\{ j=1, ..., N_\text{u}, j\neq i \big| \lVert z_i[k] -z_j[k] \rVert_2 \le d_\text{inf} \right\}
\end{align*}
for some distance $d_\text{inf}$. The sole location information in the BETIS estimation algorithm are the neighbourhoods $\mathcal{N}_{\text{u}, i}[k]$. We do not distinguish between neighbourhoods $\mathcal{N}_{\text{u}, i}[k]$ that were measured directly, by Bluetooth, or indirectly, by GPS coordinates. For the individuals $i=N_\text{u}+1, ..., N$, who do not use the app, neither location information nor health self-reports are available. Since location information for non-users $i=N_\text{u}+1, ..., N$ is not available, non-users are not registered in the neighbourhood $\mathcal{N}_{\text{u},j}[k]$ of a user $j=1, ..., N_\text{u}$. The complete neighbourhood of an individual $i$, consisting of both users and non-users, is denoted by
\begin{align}\label{N_all_from_d_inf}
\mathcal{N}_i[k] =\left\{ j=1, ..., N, j\neq i \big| \lVert z_i[k] -z_j[k] \rVert_2 \le d_\text{inf} \right\}.
\end{align}
In contrast to the neighbourhood $\mathcal{N}_{\text{u}, i}[k]$ of users, the neighbourhood $\mathcal{N}_i[k]$ is not measured. The number of contacts with non-users is denoted by 
\begin{align*}
N_{\text{nonuser},i}[k]=\left| \mathcal{N}_i[k]\right| - \left| \mathcal{N}_{\text{u}, i}[k]\right|.
\end{align*}
We assume that the distribution of the number of neighbours $N_{\text{nonuser},i}[k]$,
\begin{align*}
f(m)=\operatorname{E}_{i,k}\left[\operatorname{Pr} \left[N_{\text{nonuser},i}[k] = m\right]\right],
\end{align*}
is known, where the expectation computed with respect to every user $i$ and all times $k$. The average distribution $f(m)$ of contacts with non-users can be obtained from a representative subgroup of the population.

We model the spread of COVID-19 by a hidden Markov model, which consists of two parts. First, the dynamics of the viral state $X_i[k]$. Second, the user behaviour of reporting their viral state $X_i[k]$.

\subsection{Dynamics of the viral state $X_i[k]$}\label{subsec:dynamics}
Consider the infection of a susceptible individual $i$, with $X_i[k]=\mathcal{S}$ or $X_i[k]=\mathcal{S}_\text{fa}$. Then, individual~$i$ traverses the viral states $\mathcal{E} \rightarrow \mathcal{I}\rightarrow \mathcal{R}$ for a \textit{symptomatic} infection. Analogously, the course of an \textit{asymptomatic} infection is $\mathcal{E} \rightarrow \mathcal{I}_\text{a} \rightarrow \mathcal{R}$. The dynamics of the hidden Markov model are determined by the transition probabilities between the viral states. A susceptible individual $i$ without symptoms, $X_i[k]=\mathcal{S}$, contracts a disease with similar symptoms to COVID-19 with the probability $\vartheta$,
\begin{align*}
\operatorname{Pr}\left[ X_i[k+1] = \mathcal{S}_\text{fa} \big| X_i[k] = \mathcal{S} \right] = \vartheta,
\end{align*}
and cures with the curing probability $\delta$,
\begin{align*} 
\operatorname{Pr}\left[ X_i[k+1] = \mathcal{S} \big| X_i[k] = \mathcal{S}_\text{fa} \right] = \delta.
\end{align*}
 An infectious individual $j$, with $X_j[k]=\mathcal{I}$ or $X_j[k]=\mathcal{I}_\text{a}$, infects a susceptible individual $i$ with the infection probability $\beta$, if individual $j$ is in the neighbourhood $\mathcal{N}_i[k]$ of individual $i$. The infection probability $\beta$ depends on the contagiousness of SARS-CoV-2 and on the prevalence of facemasks and other spread reduction measures. The set
\begin{align*}
\mathcal{N}_{\text{inf},i}[k] = \left\{ j\in \mathcal{N}_i[k]  \big| X_j[k]=\mathcal{I} ~ \text{or} ~ X_j[k]=\mathcal{I}_\text{a}\right\}
\end{align*}
consists of all \textit{infectious} individuals $j$, users and non-users, that are close to individual $i$ at time $k$. The number of infectious neighbours of individual $i$ at time $k$ is denoted by $\left|\mathcal{N}_{\text{inf},i}[k] \right|$. The probability of an infection of individual $i$ follows from potential infections by any individual $j$ in the set $\mathcal{N}_{\text{inf},i}[k]$ as
\begin{align}\label{prob_E_given_S}
\operatorname{Pr}\left[ X_i[k+1] = \mathcal{E} \big| X_i[k] \in \{ \mathcal{S}, \mathcal{S}_\text{fa}\}, \mathcal{N}_{\text{inf},i}[k] \right] = 1 - \left( 1 - \beta \right)^{\left|\mathcal{N}_{\text{inf},i}[k] \right|} .
\end{align}
Individuals leave the exposed state $\mathcal{E}$ with the incubation probability~$\gamma$ to an infectious state,
\begin{align*}
\operatorname{Pr}\left[ X_i[k+1] = c \big| X_i[k] = \mathcal{E} \right]= \begin{cases}
\gamma \alpha \quad & \text{if} \quad c=\mathcal{I}_\text{a}, \\
\gamma \left(1-\alpha \right) & \text{if} \quad c=\mathcal{I}, \\
\left(1 - \gamma \right) &\text{if} \quad c=\mathcal{E}.
\end{cases}
\end{align*}
Here, $\alpha$ denotes the probability of an asymptomatic infection. Any symptomatic infected individual is removed with the removal probability $\delta$. In other words, 
\begin{align} \label{prob_curing}
\operatorname{Pr}\left[ X_i[k+1] = \mathcal{R} \big| X_i[k] = \mathcal{I} \right] = \delta.
\end{align}
 Denote the first time that individual $i$ is infected by $k_{\mathcal{I}, i}$, $X_i\left[k_{\mathcal{I}, i}\right] = \mathcal{I}$ and $X_i\left[k_{\mathcal{I}, i}-1\right] = \mathcal{E}$. Similarly, denote the first time that individual $i$ is removed by $k_{\mathcal{R}, i}$. Since the viral state compartments are in the order $\mathcal{E}\rightarrow \mathcal{I}\rightarrow \mathcal{R}$, it holds that $k_{\mathcal{R}, i} > k_{\mathcal{I}, i}$. The \textit{sojourn time} $k_{\mathcal{R}, i} - k_{\mathcal{I}, i}$ of state $\mathcal{I}$ is the number of discrete times $k$ that individual $i$ has been infected. By (\ref{prob_curing}), we implicitly assume that the sojourn time follows a geometric distribution with mean $1/\delta$.

\subsection{Reporting the viral state $X_i[k]$} \label{subsec:observations}
 If a user experiences COVID-19 related symptoms at time $k$, then the users submits a health report. We denote the \textit{reported} viral state of user $i$ as $X_{\text{rep},i}[k]$. Since the users themselves report their health status, the reported viral state $X_{\text{rep},i}[k]$ might be inaccurate. At every time $k$, the reported state $X_{\text{rep},i}[k]$ equals either: healthy $\mathcal{S}$; contracted a disease other than COVID-19, $\mathcal{S}_\text{fa}$; or infected by COVID-19, $\mathcal{I}$. A user $i$ without symptoms, $X_i[k]\in\{ \mathcal{S},\mathcal{E},\mathcal{I}_\text{a}, \mathcal{R}\}$, reports a healthy viral state $X_{\text{rep},i}[k]=\mathcal{S}$. Thus, BETIS considers that asymptomatic infections in $\mathcal{I}_\text{a}$ cannot be detected by self-reports. If user $i$ experiences symptoms that are related to COVID-19, $X_i[k]=\mathcal{S}_\text{fa}$ or $X_i[k]=\mathcal{I}$, then user $i$ specifies the symptoms via a health report in the app. Based on the health report, a user $i$ with symptoms is classified either as suffering from COVID-19, $X_{\text{rep},i}[k]=\mathcal{I}$, or from another disease, $X_{\text{rep},i}[k]=\mathcal{S}_\text{fa}$. Since the symptoms of COVID-19 overlap with symptoms of other diseases, the reported viral states $X_{\text{rep},i}[k]=\mathcal{I}$ and $X_{\text{rep},i}[k]=\mathcal{S}_\text{fa}$ can be erroneous. The errors in the reported viral state $X_{\text{rep},i}[k]$ are described by the test statistics
\begin{align*} 
\operatorname{Pr}\left[ X_{\text{rep},i}[k] = c \big| X_i[k] = \mathcal{S}_\text{fa} \right] = 
\begin{cases}
p_\text{fa} \quad &\text{if} \quad c=\mathcal{I},\\
1-p_\text{fa}&\text{if} \quad c=\mathcal{S}_\text{fa},
\end{cases}
\end{align*}
and
\begin{align*} 
\operatorname{Pr}\left[ X_{\text{rep},i}[k] = c \big| X_i[k] = \mathcal{I} \right] = 
\begin{cases}
p_\text{tp} \quad &\text{if} \quad c=\mathcal{I},\\
1-p_\text{tp}&\text{if} \quad c=\mathcal{S}_\text{fa}.
\end{cases}
\end{align*}
Hence, the accuracy of the health report is given by\footnote{In the terminology of medical testing \cite{trevethan2017sensitivity}, the probabilities $p_\text{tp}$ and $(1-p_\text{fa})$ are known as \textit{sensitivity} and \textit{specificity}, respectively.} the \textit{false alarm probability} $p_\text{fa}$ and the \textit{true positive rate} $p_\text{tp}$.

\section{Who is infected?}\label{sec:who_is_infected}
At time $k$, we would like to know who is infected by COVID-19. In other words, for every user $i$, BETIS computes the \textit{symptomatic infection risk}
\begin{align}\label{cond_prob}
\operatorname{Pr}\left[ X_i[k] = \mathcal{I} \left| \mathcal{M}[k] \right. \right]
\end{align}
and the \textit{asymptomatic infection risk}
\begin{align*}
\operatorname{Pr}\left[ X_i[k] = \mathcal{I}_\text{a} \left| \mathcal{M}[k] \right. \right].
\end{align*}
Here, we formally define all observations, or measurements, up until time $k$ as $\mathcal{M}[k]$. More specifically, the set $\mathcal{M}[k]$ specifies the reported viral state $X_{\text{rep}, i}[l]$ and the measured neighbourhood $\mathcal{N}_{\text{u},i}[l]$ of every user $i=1, ..., N_\text{u}$ at every time $l\le k$. In Appendix~\ref{appendix:computation}, we propose a recursive Bayesian filtering method to (approximately) compute the infection risks $\operatorname{Pr}\left[ X_i[k] = \mathcal{I} \left| \mathcal{M}[k] \right. \right]$ and $\operatorname{Pr}\left[ X_i[k] = \mathcal{I}_\text{a} \left| \mathcal{M}[k] \right. \right]$. As a side product, we obtain the probabilities $\operatorname{Pr}\left[ X_i[k] = c \left| \mathcal{M}[k] \right. \right]$ for the other viral states $c=\mathcal{S}, \mathcal{S}_\text{fa}, \mathcal{E}, \mathcal{R}$. The computation time is polynomial in the number of individuals $N$ and the number of observations~$k$.

We perform simulations of the epidemic model (Section~\ref{sec:model}) with $N=10,000$ moving individuals and vary the fraction of app users $c_0$. The false alarm probability is set to $p_\text{fa} =0.1$ and the positive rate to $p_\text{tp}=0.9$. \textit{We assume that none of the initial viral states $X_1[1], ..., X_N[1]$ is known to the BETIS estimation method.} Instead, we solely assume that the prior distribution of the viral state $X_i[1]$ is known. For further details on the parameter settings, we refer to Appendix~\ref{appendix:parameters}.

\subsection{Tracing the number of infections}
Can BETIS estimate the evolution of the total number of infections in the population? First, we define $\mathcal{I}_\text{all}[k]$ as the \textit{true} number of individuals, users and non-users, whose viral state $X_i[k]=\mathcal{I}$. BETIS computes the infection risks $\operatorname{Pr}\left[ X_i[k] = \mathcal{I} \left| \mathcal{M}[k] \right. \right]$ of the users $i=1, ..., N_\text{u}$. Thus, we obtain an estimate of the number of infected individuals, users and non-users, as
\begin{align*}
\hat{\mathcal{I}}_\text{all}[k] = \frac{N}{N_\text{u}} \sum^{N_\text{u}}_{i=1} \operatorname{Pr}\left[ X_i[k] = \mathcal{I} \left| \mathcal{M}[k] \right. \right].
\end{align*}
For the asymptomatic infections, the quantities $\mathcal{I}_\text{a,all}[k]$ and $\hat{\mathcal{I}}_\text{a,all}[k]$ are defined analogously. 

      \begin{figure}[!ht]
         \centering
	\subfloat[{\small Symptomatic infections $\mathcal{I}_\text{all}[k]$.}]{ \includegraphics[width=0.99\textwidth]{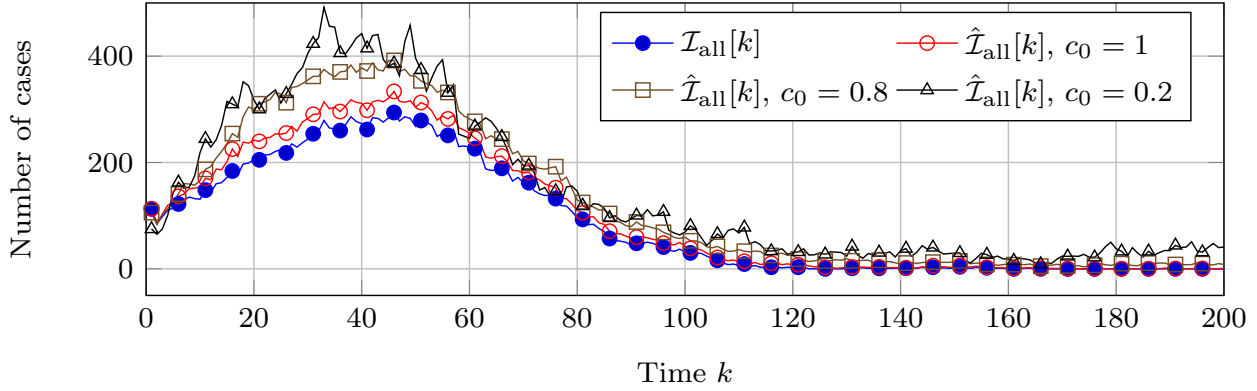}}\\
	\subfloat[{\small Asymptomatic infections $\mathcal{I}_\text{a,all}[k]$}.]{ \includegraphics[width=0.99\textwidth]{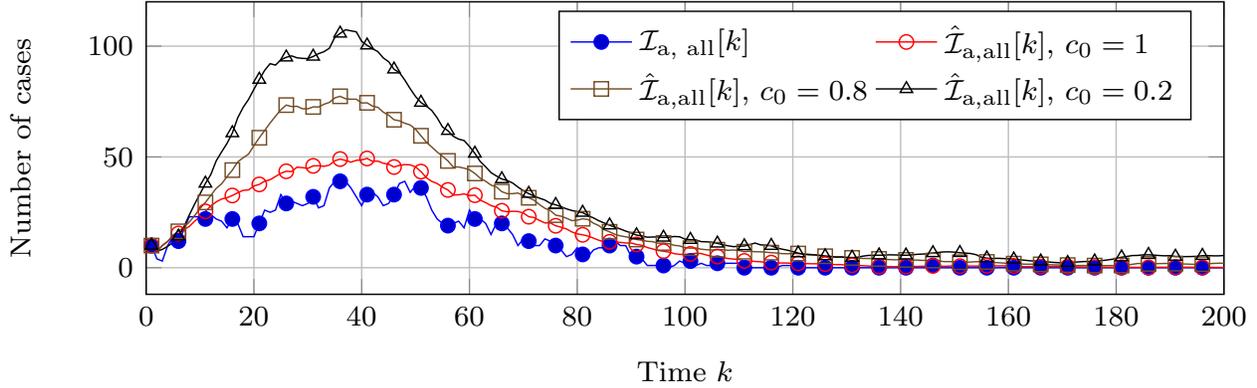}}
        \caption{\textbf{Tracing the number of infections.}The total number of symptomatic infections $\mathcal{I}_\text{all}[k]$ and asymptomatic infections $\mathcal{I}_\text{a,all}[k]$ of all individuals versus time~$k$, following the SIR epidemic model. The fraction $c_0$ of individuals, who are contact app users and report COVID-19 related symptoms, is varied. Based on the self-reports, our BETIS algorithm produces estimates $\hat{\mathcal{I}}_\text{all}[k]$, $\hat{\mathcal{I}}_\text{a,all}[k]$ for the total number of infections.}
         \label{fig:fig_1}
     \end{figure}     
     
       Figure~\ref{fig:fig_1} demonstrates the accuracy of the estimated number of symptomatic infections $\hat{\mathcal{I}}_\text{all}[k]$ and asymptomatic infections $\hat{\mathcal{I}}_\text{a,all}[k]$, for different fractions $c_0$ of individuals that use the app. Unsurprisingly, the symptomatic infections $\mathcal{I}_\text{all}[k]$ are traced more accurately than the asymptomatic infections $\mathcal{I}_\text{a,all}[k]$. For all fractions $c_0$, the simulations indicate that the BETIS estimates $\hat{\mathcal{I}}_\text{all}[k]$ and $\hat{\mathcal{I}}_\text{a,all}[k]$ are greater than\footnote{It is an open challenge to rigorously show that the BETIS overestimates the true number of infections, $\hat{\mathcal{I}}_\text{all}[k]>\mathcal{I}_\text{all}[k]$ and $\hat{\mathcal{I}}_\text{a,all}[k]>\mathcal{I}_\text{a,all}[k]$, respectively. In \cite{donnelly1993correlation, cator2014nodal} for the $N$-intertwined mean-field approximation (NIMFA) of the susceptible-infected-susceptible (SIS) epidemic process, it is shown that infection states are positively correlated, implying that an infection somewhere in the network cannot lower the probability of infection somewhere else. BETIS assumes in (\ref{assumption_1}) stochastic independence of infection states of different users, and ignoring correlations may explain the overestimations of BETIS.} the true number of infections $\mathcal{I}_\text{all}[k]$, $\mathcal{I}_\text{a, all}[k]$. From a societal point of view, overestimations give safe-side warnings, resulting in a positive property of BETIS. Overall, even if only $c_0=20\%$ individuals are users, the epidemic outbreak is traced reasonably well.       
         
       \subsection{Identifying infected individuals} 
       Beyond tracing the total number of SARS-CoV-2 infections, a tremendous challenge is to identify which users are infected. BETIS approximates the posterior probability $\operatorname{Pr}\left[ X_i[k] = c \left| \mathcal{M}[k] \right. \right]$ for every compartment $c\in\mathcal{C}$. Thus, we obtain the Bayesian estimate of the viral state $X_i[k]$ at any time $k$ as
\begin{align}\label{X_hat}
\hat{X}_i[k]=\underset{c\in\mathcal{C}}{\operatorname{arg max}} ~ \operatorname{Pr}\left[ X_i[k] = c \left| \mathcal{M}[k] \right. \right].
\end{align}
At any time $k$, the number of \textit{true positive} estimates of symptomatic infections equals the number of users $i$ for which both $X_i[k]=\mathcal{I}$ and $\hat{X}_i[k]=\mathcal{I}$. Similarly, the \textit{false positive} estimates equals the number of users $i$ for which $X_i[k]\neq\mathcal{I}$ but $\hat{X}_i[k]=\mathcal{I}$. The number of true and false positive estimates for asymptomatic infections is defined analogously.

      \begin{figure}[!ht]
         \centering
	\subfloat[{\small Symptomatic infections.}]{ \includegraphics[width=0.99\textwidth]{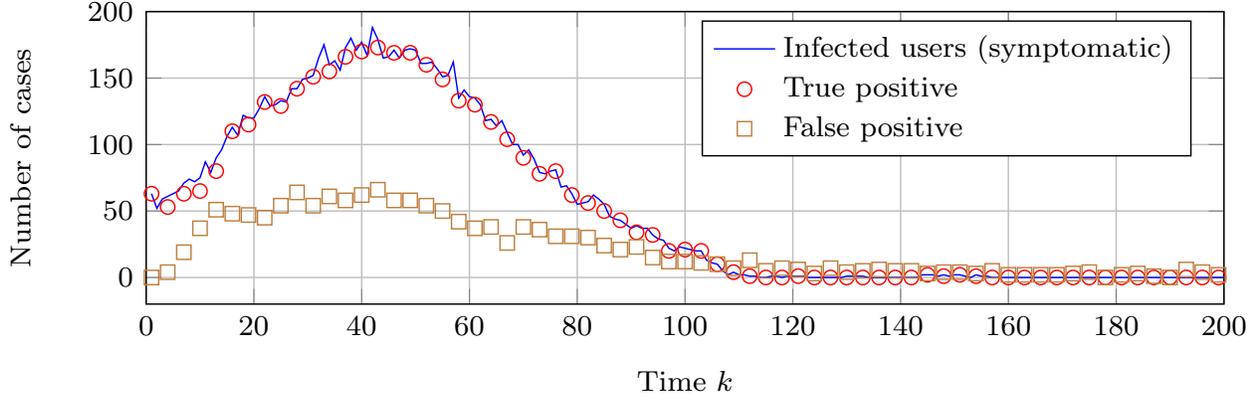}}\\
	\subfloat[{\small Asymptomatic infections.}]{ \includegraphics[width=0.99\textwidth]{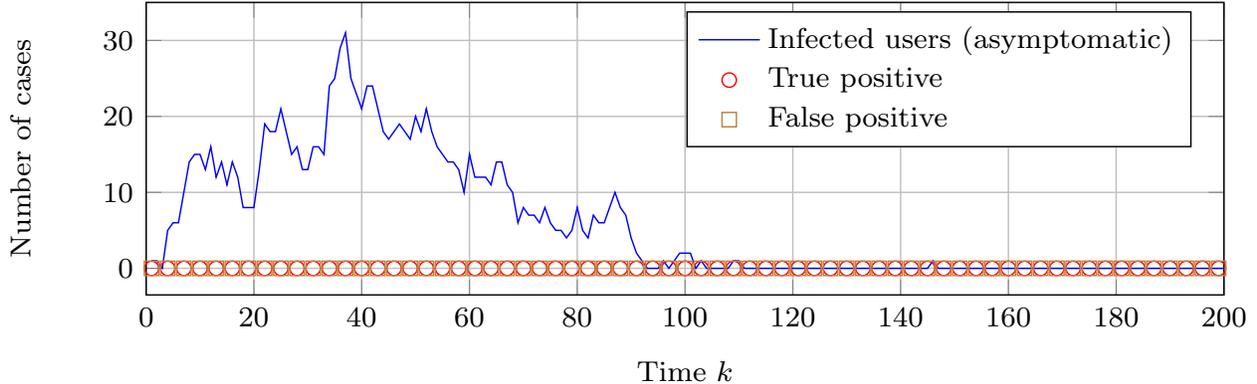}}
        \caption{\textbf{Identifying infected users.} The solid line in the subplots depicts the number of users with symptomatic and asymptomatic infections, respectively. The marks correspond to the number of users that BETIS correctly (true positive) and incorrectly (false positive) identifies as infectious.}
         \label{fig:fig_2}
     \end{figure}   
     
     In the following, we assume that a fraction of $c_0=0.6$ individuals use the app. Figure~\ref{fig:fig_2} demonstrates the accuracy of identifying infectious individuals by the BETIS estimation algorithm. BETIS performs well for identifying symptomatic infections: Almost every symptomatic infection is correctly identified (true positives), with relatively few false positives. On the other hand, Figure~\ref{fig:fig_2} shows that \textit{asymptomatic} infections cannot be directly identified by (\ref{X_hat}): There is no user $i$ whose most likely state is asymptomatic infectious $\hat{X}_i[k]=\mathcal{I}_\text{a}$, which contrasts the accuracy of BETIS in tracing the \textit{total} number of asymptomatic infections $\mathcal{I}_\text{a, all}[k]$, see Figure~\ref{fig:fig_1}. 
     
\begin{figure}[!ht]
         \centering
         \includegraphics[width=\textwidth]{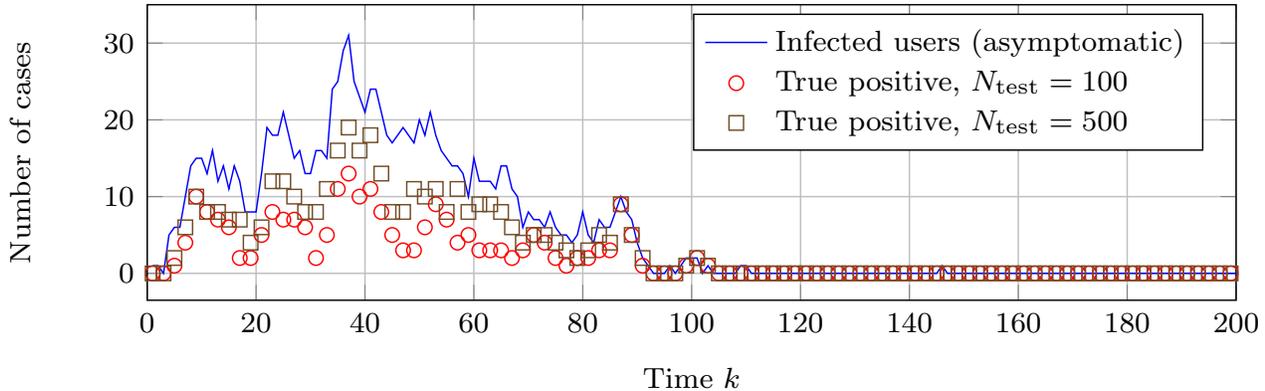}
         \caption{\textbf{Testing users without symptoms to identify asymptomatic infections.} The solid line depicts the number of users with an asymptomatic infection, where the total number of users equals $N_\text{u}=6,000$. To detect asymptomatic infections, a limited number of $N_\text{test}$ users without symptoms are tested for COVID-19. These $N_\text{test}$ tested users have the highest risk of an asymptomatic infection, as computed by BETIS. The marks correspond to the number of positive test results.\label{fig:fig_3}}
     \end{figure}

Nonetheless, we show in Figure~\ref{fig:fig_3} that BETIS is valuable for identifying asymptomatic infections. Health agencies rely on reverse transcription polymerase chain reaction (RT-PCR) test methods to accurately determine whether an individual is infected by SARS-CoV-2. In an ideal, utopian scenario, there would be sufficient RT-PCR testing capacities to check every individual regularly, such that every asymptomatic infection would be detected timely. However, the testing capacities are insufficient, and only a limited number of people can be tested by RT-PCR methods. Specifically, suppose that only $N_\text{test}<<N$ individuals can be tested for the identifying asymptomatic infections. Which $N_\text{test}$ individuals are the most likely to suffer from an asymptomatic infection and return a positive test result? Our approach is to select those $N_\text{test}$ users who have the greatest probability of an asymptomatic infection, $\operatorname{Pr}\left[ X_i[k] = \mathcal{I}_\text{a} \left| \mathcal{M}[k] \right. \right]$, which is computed by BETIS.

     Figure~\ref{fig:fig_3} shows that the contact app indeed helps in identifying users with asymptomatic infections. We emphasise that $N_\text{test}=100$ tests corresponds to testing less than $2\%$ of the users. Furthermore, \textit{group testing} methods \cite{du2000combinatorial} are able to identify all infections within a group of $N_\text{test}$ individuals, by using significantly less than $N_\text{test}$ tests. In particular, the combination of the group testing method for SARS-CoV-2 by Shental \textit{et al.} \cite{shental2020efficient} with BETIS is a promising approach to detect the majority of asymptomatic users.  
     
     \subsection{Performance limits}
                       
      \begin{figure}[!ht]
         \centering
	\subfloat[{\small Symptomatic infections $\mathcal{I}_\text{all}[k]$.}]{ \includegraphics[width=0.99\textwidth]{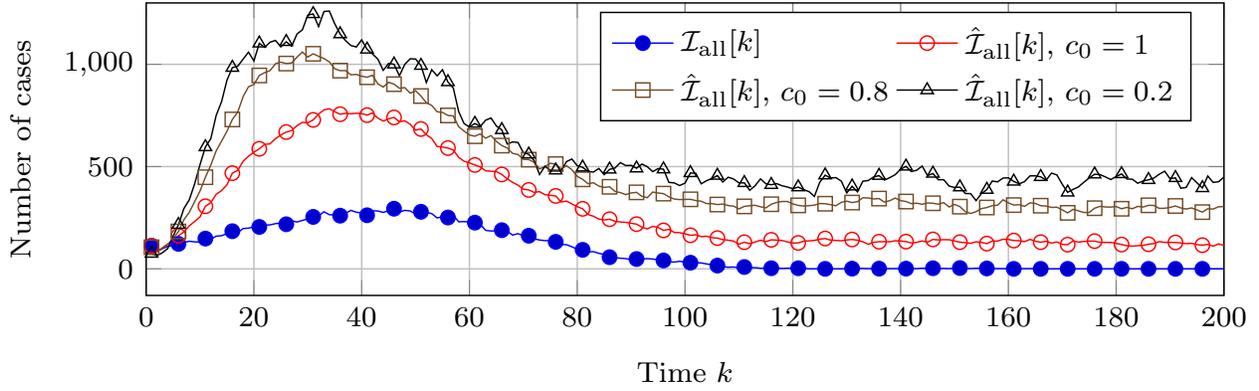}}\\
	\subfloat[{\small Asymptomatic infections $\mathcal{I}_\text{a,all}[k]$}.]{ \includegraphics[width=0.99\textwidth]{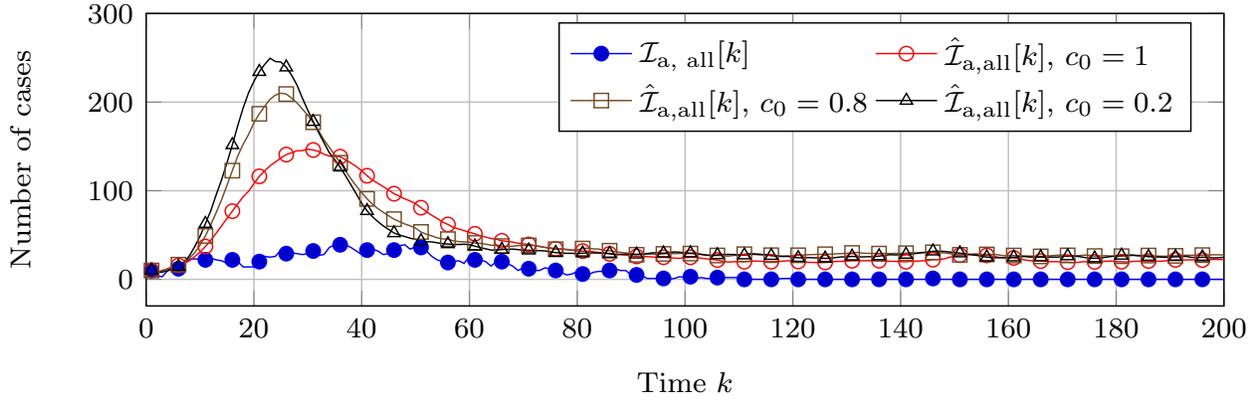}}
        \caption{\textbf{Tracing the number of infections (inaccurate health reports).} The total number of symptomatic infections $\mathcal{I}_\text{all}[k]$, asymptomatic infections $\mathcal{I}_\text{a,all}[k]$ and the respective BETIS estimates $\hat{\mathcal{I}}_\text{all}[k]$, $\hat{\mathcal{I}}_\text{a,all}[k]$. In comparison to Figure~\ref{fig:fig_1}, the health report by the users is less reliable.}
         \label{fig:fig_1_limits}
     \end{figure}
     
          \begin{figure}[!ht]
         \centering
         \includegraphics[width=\textwidth]{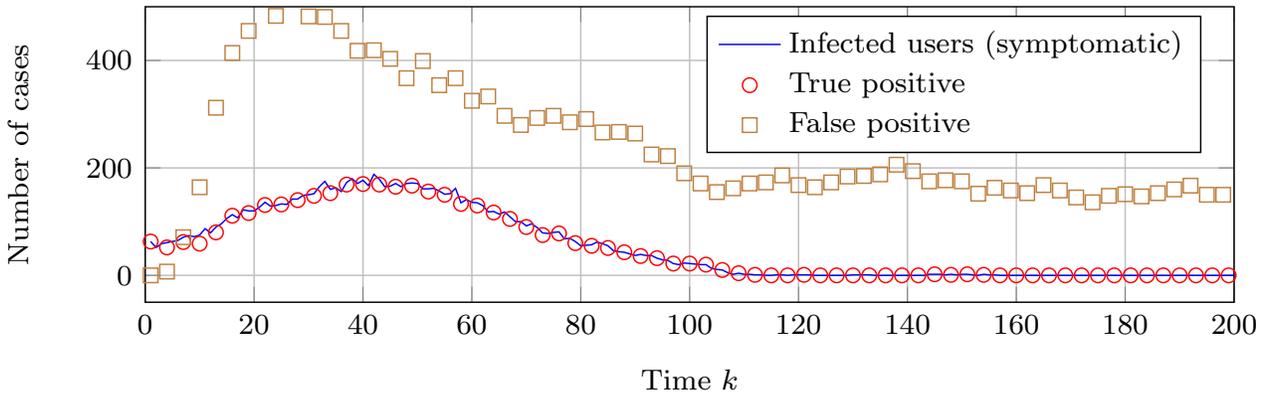}
         \caption{\textbf{Identifying infected users (inaccurate health reports).}  The solid line depicts the number of users with a symptomatic infection. The marks correspond to the number of users that BETIS correctly (true positive) and incorrectly (false positive) identifies as infectious. In comparison to Figure~\ref{fig:fig_2}, the health report by the users is less reliable.}   \label{fig:fig_2_I_limits}
     \end{figure}
     
           \begin{figure}[!ht]
         \centering
         \includegraphics[width=\textwidth]{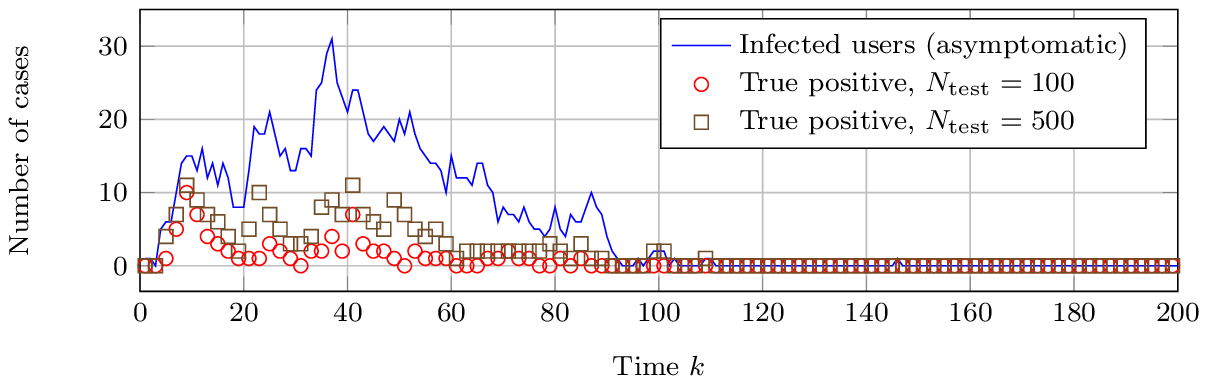}
         \caption{\textbf{Testing users without symptoms to identify asymptomatic infections (inaccurate health reports).} The solid line depicts the number of users with an asymptomatic infection, where the total number of users equals $N_\text{u}=6,000$. The marks correspond to the number of positively tested users, when $N_\text{test}$ users are tested. In comparison to Figure~\ref{fig:fig_3}, the health report by the users is less reliable.}
         \label{fig:fig_3_limits}
     \end{figure}
     
The value of BETIS lies in jointly processing the location information and health reports of the users. Thus, the accuracy of BETIS depends on the testing statistic of the self-reports. We deteriorate the test statistics by increasing the false alarm probability to $p_\text{fa} =0.2$ and decreasing the true positive rate to $p_\text{tp}=0.75$.
        
    Figures~\ref{fig:fig_1_limits}--\ref{fig:fig_3_limits}, in comparison with Figures~\ref{fig:fig_1}--\ref{fig:fig_3}, shows that inaccurate health reports directly affect the accuracy of tracing the number of infections and identifying infectious users. Hence, the development of accurate methods for assessing the user's health status are important. Nonetheless, even for inaccurate health reports, BETIS yields a valuable upper bound of the number of infections and helps in identifying both symptomatic and asymptomatic infections.

\section{Conclusions}
This work considers the application of contact apps beyond alarming users of potential infections: the \textit{detection} of SARS-CoV-2 infections. The app tracks the location of the users, and inquires self-reports of the user's health status. No information is required on individuals that do not use the app. 

We propose the BETIS algorithm for detecting infections, based on the measurements of the contact app. BETIS detects the SARS-CoV-2 infections of every user within a reasonable accuracy, even if only a fraction of the population use the contact app. Furthermore, in spite of many uncertainties, BETIS operates on the safe-side of detection by surprisingly accurately overestimating infected individuals. BETIS thus constitutes a major tool for detecting infections in any pandemic.

We emphasise that there is a twofold benefit for every person who installs the app. First, every single user actively contributes to tracing and eradicating SARS-CoV-2, which is advantageous to the whole society. Second, there is an immediate personal benefit for every app user: am I infected or not? The combination of contributing to society and gaining information on the personal health is a great incentive to install the app.

 The algorithmic framework of BETIS can be used as basis for further improvements. Of particular interest are human mobility patterns, to obtain a more accurate estimate of the interactions between users and non-users. Another direction is the use of measurements additional to the health self reports, such as randomised COVID-19 tests of the whole population.

\section*{Acknowledgements}
This work has been supported by the Universiteitsfonds Delft in the program TU Delft COVID-19 Response Fund.

\appendix

\section{The BETIS algorithm}\label{appendix:computation}

\subsection{Assumptions in the computations}\label{subsec:assumptions}
 We define the $N_\text{u}\times 1$ viral state vector as $X[k]= \left(X_1[k], ..., X_{N_\text{u}}[k]\right)^T$. The reported viral state vector $X_{\text{rep}}[k]$ is defined analogously. We rely on three assumptions to compute the infection risk (\ref{cond_prob}). First, we assume the conditional stochastic independence
 \begin{align}\label{assumption_1}
\operatorname{Pr}\left[ X[k]\big| X_{\text{rep}}[k], \mathcal{M}[k-1] \right]
\approx \prod^{N_\text{u}}_{i=1} \operatorname{Pr}\left[ X_i[k]\big| X_{\text{rep}, i}[k], \mathcal{M}[k-1] \right].
 \end{align}
There are $6^{N_\text{u}}$ possible combinations of the entries of the viral state vector $X[k]$. Thus, it is practically impossible to state the full distribution of the vector $X[k]$. The assumption (\ref{assumption_1}) instead implies that the distribution of the vector $X[k]$ can be decomposed into the marginal distribution of the entries $X_1[k]$, $X_2[k]$, ..., $X_{N_\text{u}}[k]$, which can be computed separately. Furthermore, assumption (\ref{assumption_1}) might be of relevance to privacy: The full distribution $\operatorname{Pr}\left[ X[k]\big| X_{\text{rep}}[k], \mathcal{M}[k-1] \right]$ is sensitive data. In contrast, the single factors $\operatorname{Pr}\left[ X_i[k]\big| X_{\text{rep}, i}[k], \mathcal{M}[k-1] \right]$ might \textit{in parts} be made accessible to some individuals. 

Furthermore, we make the assumption that the viral state $X_i[k]$ does not depend on the measured neighbourhoods $\mathcal{N}_{\text{u},1}[k], ..., \mathcal{N}_{\text{u},N_\text{u}}[k]$ at time $k$. More precisely,
\begin{align}\label{assumption_2}
\operatorname{Pr}\left[ X_i[k]\big| X_{\text{rep}, i}[k], \mathcal{N}_{\text{u},1}[k], ...,  \mathcal{N}_{\text{u},N_\text{u}}[k],\mathcal{M}[k-1] \right]
= \operatorname{Pr}\left[ X_i[k]\big| X_{\text{rep}, i}[k], \mathcal{M}[k-1] \right].
\end{align} 
The viral state $X_i[k]$ does depend on the neighbourhoods $\mathcal{N}_{\text{u},i}[k-1]$ at the \textit{previous} time step $k-1$, due to the infection probability (\ref{prob_E_given_S}). Thus, the impact of the location on the infection dynamics is delayed by one time step, and we consider assumption (\ref{assumption_2}) rather technical. Third, we assume the analogue to (\ref{assumption_2}) for the \textit{joint} distribution of the random variables $X_1[k], ..., X_{N_\text{u}}[k]$,
\begin{align}\label{assumption_3}
\operatorname{Pr}\left[X[k] \big| X_{\text{rep}}[k],\mathcal{N}_{\text{u},1}[k], ..., \mathcal{N}_{\text{u},N_\text{u}}[k], \mathcal{M}[k-1] \right] =\operatorname{Pr}\left[X[k] \big| X_{\text{rep}}[k], \mathcal{M}[k-1] \right] .
\end{align} 

\subsection{Approximation of the infection probability}\label{subsec:apx_inf_prob}
BETIS computes the infection risk (\ref{cond_prob}) based on the hidden Markov epidemic model in Section~\ref{sec:model}. However, the location of non-users is unknown. Hence, the set $\mathcal{N}_{\text{inf},i}[k]$ of infectious neighbours is not known, and the infection probability (\ref{prob_E_given_S}) cannot be computed directly. Instead, we resort to approximating the infection probability (\ref{prob_E_given_S}), based on the neighbourhood of infected \textit{users} as
\begin{align*}
\mathcal{N}_{\text{inf},\text{u},i}[k] = \left\{ j\in \mathcal{N}_{\text{u},i}[k]  \big| X_j[k]=\mathcal{I} ~ \text{or} ~ X_j[k]=\mathcal{I}_\text{a}\right\}.
\end{align*}
In contrast to the complete infectious neighbourhood $\mathcal{N}_{\text{inf},i}[k]$, the subset $\mathcal{N}_{\text{inf},\text{u},i}[k]$ can be inferred from the measured neighbourhood $\mathcal{N}_{\text{u},i}[k]$, as detailed in Subsection~\ref{subsec:recursive_filtering}.

With the set $\mathcal{N}_{\text{inf},\text{u},i}[k]$, we approximate the infection probability (\ref{prob_E_given_S}) in two steps. First, at any time $k$, we approximate the probability that a randomly chosen non-user is infected (symptomatically or asymptomatically) by averaging over the infection probability of the users as
\begin{align*}
p_\text{inf}[k] = \frac{1}{N_\text{u}}\sum^{N_\text{u}}_{i=1} \left( \operatorname{Pr}\left[ X_i[k] = \mathcal{I} \left| \mathcal{M}[k] \right. \right] + \operatorname{Pr}\left[ X_i[k] = \mathcal{I}_\text{a} \left| \mathcal{M}[k] \right. \right] \right).
\end{align*}
Then, the probability that, out of $m$ randomly chosen non-users, $l$ individuals are infected follows as 
\begin{align*}
p_{\text{inf},l}[k] = {m \choose l} p^l_\text{inf}[k] \left( 1 - p_\text{inf}[k]  \right)^{m-l}. 
\end{align*} 
Thus, given that a user has $m$ contacts with non-users, the probability of an infection by a non-user equals
\begin{align*}
\epsilon[ k,m] = \sum^m_{l=0} p_{\text{inf},l}[k] \left( 1 - \left( 1 - \beta \right)^l \right).
\end{align*}
The distribution $f(m)$ of the number of contacts with non-users is known. Hence, the probability that a user is infected by a non-user is approximated by
\begin{align}\label{kjnkjnikjn}
\epsilon[k] = \sum^\infty_{m=0} f(m)  \epsilon[ k,m].
\end{align}
Second, we use (\ref{kjnkjnikjn}) to approximate the infection probability (\ref{prob_E_given_S}). More precisely, BETIS replaces the exact probability (\ref{prob_E_given_S}) by
\begin{align}\label{prob_E_given_S_apx}
\operatorname{Pr}\left[ X_i[k+1]=\mathcal{E}\big| X_i[k]\in \{ \mathcal{S}, \mathcal{S}_\text{fa}\},\left| \mathcal{N}_{\text{inf},\text{u},i}[k]\right|\right] \approx 1- \left( 1 - \beta \right)^{\left| \mathcal{N}_{\text{inf},\text{u},i}[k]\right|} \left( 1 - \epsilon[k]\right).
\end{align}

\subsection{Recursive Bayesian filtering} \label{subsec:recursive_filtering}
The infection risk (\ref{cond_prob}) can be computed by iterating over time:
\begin{description}
\item[Initialisation] At time $k=1$, we assume that the probability distribution
\begin{align*}
\operatorname{Pr}\left[ X_i[1] \right] 
\end{align*}
is given for every user $i=1, ..., N_\text{u}$. Formally, we can write
\begin{align}\label{asdasfsdfds}
\operatorname{Pr}\left[ X_i[1] \right]  = \operatorname{Pr}\left[ X_i[1] \big| \mathcal{M}[0]\right],
\end{align}
since there are no observations at time $k=0$. (Or, the set of observation $\mathcal{M}[0]$ at time $k=0$ is empty, because we start measuring at $k=1$.)

\item[Measurement update] We are given the distribution $\operatorname{Pr}\left[ X_i[k]\big| \mathcal{M}[k-1] \right]$ for every user~$i$. (Starting with (\ref{asdasfsdfds}) at time $k=1$.) For every user $i$, the measurement update incorporates the reported viral state $X_{\text{rep},i}[k]$ to obtain a more accurate distribution of the viral state $X_i[k]$. More precisely, we compute the probability $\operatorname{Pr}\left[ X_i[k]\big| X_{\text{rep}, i}[k], \mathcal{M}[k-1] \right]$ with Bayes' Theorem \cite{van2014performance} as
\begin{align*}
\operatorname{Pr}\left[ X_i[k]\big| X_{\text{rep},i}[k], \mathcal{M}[k-1] \right] &=\frac{\operatorname{Pr}\left[ X_{\text{rep},i}[k] \big| X_i[k], \mathcal{M}[k-1] \right] \operatorname{Pr}\left[ X_i[k] \big| \mathcal{M}[k-1] \right]}{\operatorname{Pr}\left[  X_{\text{rep},i}[k] \big| \mathcal{M}[k-1] \right] }.
\end{align*}
Given the viral state $X_i[k]$, the reported viral state $X_{\text{rep},i}[k]$ does not depend on past measurements $\mathcal{M}[k-1]$, and hence
\begin{align} \label{asdsdffds}
\operatorname{Pr}\left[ X_i[k]\big| X_{\text{rep},i}[k], \mathcal{M}[k-1] \right] &=\frac{\operatorname{Pr}\left[ X_{\text{rep},i}[k] \big| X_i[k] \right] \operatorname{Pr}\left[ X_i[k] \big| \mathcal{M}[k-1] \right]}{\operatorname{Pr}\left[  X_{\text{rep},i}[k] \big| \mathcal{M}[k-1] \right] }.
\end{align}
The distribution $\operatorname{Pr}\left[ X_{\text{rep},i}[k] \big| X_i[k] \right]$ is specified by the observation model in Subsection~\ref{subsec:observations}. In particular, for $X_{\text{rep}, i}[k]=\mathcal{R}$, it holds that 
\begin{align*}
\operatorname{Pr}\left[  X_{\text{rep}, i}[k]=\mathcal{R} \big| X_i[k]=c, \mathcal{M}[k-1] \right] = \begin{cases}
1 \quad &\text{if} \quad   c= \mathcal{R},\\
0 &\text{if} \quad   c\neq \mathcal{R}.
\end{cases}
\end{align*}
If user $i$ reports to be healthy, $X_{\text{rep}, i}[k]=\mathcal{S}$, then we obtain that
\begin{align*}
\operatorname{Pr}\left[  X_{\text{rep}, i}[k]=\mathcal{S} \big| X_i[k]=c, \mathcal{M}[k-1] \right] = \begin{cases}
1 \quad &\text{if} \quad   c\in \{ \mathcal{E}, \mathcal{I}_\text{a}, \mathcal{R}_\text{a}\},\\
1-p_\text{fa} &\text{if} \quad   c= \mathcal{S},\\
0 &\text{if} \quad   c\in \{ \mathcal{I}, \mathcal{R}\}.
\end{cases}
\end{align*}
Similarly, if user $i$ reports to be infected, $X_{\text{rep}, i}[k]=\mathcal{I}$, then it holds that
\begin{align*}
\operatorname{Pr}\left[  X_{\text{rep}, i}[k]=\mathcal{I} \big| X_i[k]=c, \mathcal{M}[k-1] \right] = \begin{cases}
1 \quad &\text{if} \quad   c= \mathcal{I},\\
p_\text{fa} &\text{if} \quad   c= \mathcal{S},\\
0 &\text{if} \quad   c \in \{ \mathcal{E}, \mathcal{R}, \mathcal{I}_\text{a}, \mathcal{R}_\text{a}\}.
\end{cases}
\end{align*}
The denominator in (\ref{asdsdffds}) follows from the law of total probability \cite{van2014performance} as
\begin{align*}
\operatorname{Pr}\left[  X_{\text{rep},i}[k] \big| \mathcal{M}[k-1] \right] = \sum_{c\in \mathcal{C}} \operatorname{Pr}\left[  X_{\text{rep},i}[k] \big| X_i[k]=c\right] \operatorname{Pr}\left[  X_i[k]=c \big| \mathcal{M}[k-1]\right].
\end{align*}

\item[Time update] The measurement update computes the distribution $\operatorname{Pr}\left[ X_i[k]\big| X_{\text{rep},i}[k], \mathcal{M}[k-1] \right]$, from which the time update obtains the distribution $\operatorname{Pr}\left[ X_i[k+1]\big| \mathcal{M}[k] \right]$. The law of total probability yields that
\begin{align}
\operatorname{Pr}\left[ X_i[k+1]\big| \mathcal{M}[k] \right] &= \sum_{c\in \mathcal{C}} \operatorname{Pr}\left[ X_i[k+1], X_i[k] = c\big| \mathcal{M}[k] \right] \nonumber\\
&= \sum_{c\in \mathcal{C}} \operatorname{Pr}\left[ X_i[k+1]\big| X_i[k]=c, \mathcal{M}[k] \right]  \operatorname{Pr}\left[ X_i[k]=c\big| \mathcal{M}[k] \right], \label{ljnkjndsfd}
\end{align}
where the last equation follows from the definition of the conditional probability. First, we consider the term $\operatorname{Pr}\left[ X_i[k]=c\big| \mathcal{M}[k] \right]$ in (\ref{ljnkjndsfd}). With the definition of the set of all observations $\mathcal{M}[k]$, it holds that
\begin{align*}
\operatorname{Pr}\left[ X_i[k]=c\big| \mathcal{M}[k] \right] =
\operatorname{Pr}\left[ X_i[k]=c\big| X_\text{rep}[k], \mathcal{N}_{\text{u},1}[k], ..., \mathcal{N}_{\text{u},N_\text{u}}[k], \mathcal{M}[k-1] \right]. 
\end{align*}
Assumption (\ref{assumption_1}) implies that 
\begin{align*}
\operatorname{Pr}\left[ X_i[k]=c\big| \mathcal{M}[k] \right] =
\operatorname{Pr}\left[ X_i[k]=c\big| X_{\text{rep}, i}[k], \mathcal{N}_{\text{u},1}[k], ..., \mathcal{N}_{\text{u},N_\text{u}}[k], \mathcal{M}[k-1] \right]. 
\end{align*}
Then, with assumption (\ref{assumption_2}), we obtain that 
\begin{align} \label{ljnkjqww}
\operatorname{Pr}\left[ X_i[k]=c\big| \mathcal{M}[k] \right] =
\operatorname{Pr}\left[ X_i[k]=c\big| X_{\text{rep}, i}[k], \mathcal{M}[k-1] \right],
\end{align}
which has been calculated by the previous measurement update. Second, we consider the term $\operatorname{Pr}\left[ X_i[k+1]\big| X_i[k]=c, \mathcal{M}[k] \right]$ in (\ref{ljnkjndsfd}). The exact transition probabilities of the viral state $X_i[k]$ from time $k$ to $k+1$ depends on the infectious neighbourhood $\mathcal{N}_{\text{inf},i}[k]$, as specified by the Markov epidemic model. The complete neighbourhood $\mathcal{N}_{\text{inf},i}[k]$ of infectious individuals is not measured. Thus, BETIS makes use of the transition probability approximation (\ref{prob_E_given_S_apx}), which is based on the neighbourhood $\mathcal{N}_{\text{inf},\text{u},i}[k]$ of infectious users. However, we do not directly observe the set $\mathcal{N}_{\text{inf},\text{u},i}[k]$ but instead the set $\mathcal{N}_{\text{u},i}[k]$ of all users, infectious and non-infectious, that were close to user $i$ at time $k$. Since $\mathcal{N}_{\text{inf},\text{u},i}[k] \subset \mathcal{N}_{\text{u},i}[k]$, it holds that 
\begin{align*}
0 \le \left| \mathcal{N}_{\text{inf},\text{u},i}[k]\right| \le \left| \mathcal{N}_{\text{u},i}[k]\right|.
\end{align*}
Thus, we can apply the law of total probability to obtain that 
\begin{align*}
\operatorname{Pr}\left[ X_i[k+1]\big| X_i[k]=c, \mathcal{M}[k] \right] = \sum^{\left| \mathcal{N}_{\text{u},i}[k]\right|}_{m = 0}  &\operatorname{Pr}\left[ X_i[k+1]\big| X_i[k]=c, \mathcal{M}[k],\left| \mathcal{N}_{\text{inf},\text{u},i}[k]\right|=m\right] \\
&\cdot \operatorname{Pr}\left[ \left| \mathcal{N}_{\text{inf},\text{u},i}[k]\right|=m \big| X_i[k]=c, \mathcal{M}[k] \right],
\end{align*}
which simplifies to 
\begin{align}\label{asdasddfddd}
\operatorname{Pr}\left[ X_i[k+1]\big| X_i[k]=c, \mathcal{M}[k] \right] = \sum^{\left| \mathcal{N}_{\text{u},i}[k]\right|}_{m = 0}  &\operatorname{Pr}\left[ X_i[k+1]\big| X_i[k]=c, \left| \mathcal{N}_{\text{inf},\text{u},i}[k]\right|=m\right] \\
&\cdot \operatorname{Pr}\left[ \left| \mathcal{N}_{\text{inf},\text{u},i}[k]\right|=m \big|  \mathcal{M}[k] \right].\nonumber
\end{align} 
The probabilities $\operatorname{Pr}\left[ X_i[k+1]\big| X_i[k]=c, \left| \mathcal{N}_{\text{inf},i}[k]\right|=m\right]$ are fully specified by the hidden Markov model in Subsection~\ref{subsec:dynamics} and the approximation~(\ref{prob_E_given_S_apx}). Particularly, for the susceptible compartment $X_i[k]=\mathcal{S}$, we obtain with~(\ref{prob_E_given_S_apx}) that
\begin{align*}
\operatorname{Pr}\left[ X_i[k+1]=c\big| X_i[k]=\mathcal{S},\left| \mathcal{N}_{\text{inf},\text{u},i}[k]\right|=m\right] =
\begin{cases}
\left( 1 - \beta \right)^{m}\left( 1 - \beta \right)^{\bar{\mathcal{N}}_\text{inf}} \quad &\text{if} \quad c=\mathcal{S}, \\
1 - \left( 1 - \beta \right)^{m}\left( 1 - \beta \right)^{\bar{\mathcal{N}}_\text{inf}}  &\text{if} \quad c=\mathcal{E},\\
0 &\text{otherwise}.
\end{cases}
\end{align*}
 To compute (\ref{asdasddfddd}), it remains to determine the probabilities $\operatorname{Pr}\left[ \left| \mathcal{N}_{\text{inf},\text{u},i}[k]\right|=m \big|  \mathcal{M}[k] \right]$ for all cardinalities $m=0, 1, ..., \left| \mathcal{N}_{\text{u},i}[k]\right|$. Without loss of generality\footnote{Otherwise, consider a relabelling of the nodes $j$ in the set $\mathcal{N}_{\text{u},i}[k]$.}, we assume that the neighbourhood of users~$i$ at time $k$ equals
\begin{align*}
\mathcal{N}_{\text{u},i}[k] = \{1, 2, ..., M \},
\end{align*}
where $M=\left| \mathcal{N}_{\text{u},i}[k]\right|$. The law of total probability yields that
\begin{multline*}
\operatorname{Pr}\left[ \left| \mathcal{N}_{\text{inf},\text{u},i}[k]\right|=m \big|  \mathcal{M}[k] \right] = 
\sum_{c_1\in \mathcal{C}} ... \sum_{c_M\in \mathcal{C}}\operatorname{Pr}\left[ \left| \mathcal{N}_{\text{inf},\text{u},i}[k]\right|=m \big| X_1[k]=c_1, ..., X_M[k]=c_M, \mathcal{M}[k] \right]\\
\cdot \operatorname{Pr}\left[X_1[k]=c_1, ..., X_M[k]=c_M \big| \mathcal{M}[k] \right].
\end{multline*}
 With the definition of the set of all observations $\mathcal{M}[k]$, we obtain that
\begin{multline*}
\operatorname{Pr}\left[ \left| \mathcal{N}_{\text{inf},\text{u},i}[k]\right|=m \big|  \mathcal{M}[k] \right] = 
\sum_{c_1\in \mathcal{C}} ... \sum_{c_M\in \mathcal{C}} \operatorname{Pr}\left[ \left| \mathcal{N}_{\text{inf},\text{u},i}[k]\right|=m \big| X_1[k]=c_1, ..., X_M[k]=c_M, \mathcal{M}[k] \right]\\
\cdot \operatorname{Pr}\left[X_1[k]=c_1, ..., X_M[k]=c_M \big| X_{\text{rep}}[k],\mathcal{N}_{\text{u},1}[k], ..., \mathcal{N}_{\text{u},N}[k], \mathcal{M}[k-1] \right].
\end{multline*}
The neighbourhood $\mathcal{N}_{\text{inf},\text{u},i}[k]$ of infectious users is completely determined by the viral states $X_i[k]$ of every user $i$. Thus, it holds that
\begin{multline*}
\operatorname{Pr}\left[ \left| \mathcal{N}_{\text{inf},\text{u},i}[k]\right|=m \big|  \mathcal{M}[k] \right] = 
\sum_{c_1\in \mathcal{C}} ... \sum_{c_M\in \mathcal{C}} \operatorname{Pr}\left[ \left| \mathcal{N}_{\text{inf},\text{u},i}[k]\right|=m \big| X_1[k]=c_1, ..., X_M[k]=c_M \right]\\
\cdot \operatorname{Pr}\left[X_1[k]=c_1, ..., X_M[k]=c_M \big| X_{\text{rep}}[k],\mathcal{N}_{\text{u},1}[k], ..., \mathcal{N}_{\text{u},N}[k], \mathcal{M}[k-1] \right].
\end{multline*}
From assumption (\ref{assumption_3}), it follows that
\begin{multline*}
\operatorname{Pr}\left[ \left| \mathcal{N}_{\text{inf},\text{u},i}[k]\right|=m \big|  \mathcal{M}[k] \right] = 
\sum_{c_1\in \mathcal{C}} ... \sum_{c_M\in \mathcal{C}} \operatorname{Pr}\left[ \left| \mathcal{N}_{\text{inf},\text{u},i}[k]\right|=m \big| X_1[k]=c_1, ..., X_M[k]=c_M\right]\\
\cdot \operatorname{Pr}\left[X_1[k]=c_1, ..., X_M[k]=c_M \big| X_{\text{rep}}[k], \mathcal{M}[k-1] \right].
\end{multline*}
With assumption (\ref{assumption_1}), we obtain that
\begin{multline}\label{dafkchsaakwefw}
\operatorname{Pr}\left[ \left| \mathcal{N}_{\text{inf},\text{u},i}[k]\right|=m \big|  \mathcal{M}[k] \right] = 
\sum_{c_1\in \mathcal{C}} ... \sum_{c_M\in \mathcal{C}} \operatorname{Pr}\left[ \left| \mathcal{N}_{\text{inf},\text{u},i}[k]\right|=m \big| X_1[k]=c_1, ..., X_M[k]=c_M \right]\\
\prod^M_{j=1}\operatorname{Pr}\left[X_j[k]=c_j \big| X_{\text{rep}, j}[k], \mathcal{M}[k-1] \right].
\end{multline}
The set $\mathcal{N}_{\text{inf},\text{u},i}[k]$ only consists of users $j$ with $X_j[k]=\mathcal{I}$ or $X_j[k]=\mathcal{I}_\text{a}$. For $j=1, ..., M$, we define the Bernoulli random variable $\psi_j$ as
\begin{align*} 
\psi_j = \begin{cases}
1 \quad &\text{with probability} \quad p_j, \\
0&\text{with probability} \quad 1-p_j,
\end{cases}
\end{align*}
 with the success probability
\begin{align*}
p_j = \operatorname{Pr}\left[X_j[k]=\mathcal{I} \big| X_{\text{rep}, j}[k], \mathcal{M}[k-1] \right] + \operatorname{Pr}\left[X_j[k]=\mathcal{I}_\text{a} \big| X_{\text{rep}, j}[k], \mathcal{M}[k-1] \right].
\end{align*} 
 From (\ref{dafkchsaakwefw}) it follows that the cardinality $\left| \mathcal{N}_{\text{inf},\text{u},i}[k]\right|$ is the sum of $M$ Bernoulli random variables $\psi_j\in \{0,1\}$ with different success probabilities $p_j$. Hence, the cardinality $\left| \mathcal{N}_{\text{inf},\text{u},i}[k]\right|$ follows a \textit{Poisson binomial distribution}~\cite{hong2013computing}. We obtain the distribution of $\left| \mathcal{N}_{\text{inf},\text{u},i}[k]\right|$ by convolution of the distributions of the random variables $\psi_1,..., \psi_M$. If the number $M$ is large, then the convolution might take long. For large $M$, there are more efficient algorithms \cite{hong2013computing} for computing the distribution of the cardinality $\left| \mathcal{N}_{\text{inf},\text{u},i}[k]\right|$ (based on the discrete Fourier transform).
 
\end{description}
After the initialisation, the measurement update and the time update are alternated for every time~$k$. Finally, the risk factor (\ref{cond_prob}) is obtained from (\ref{ljnkjqww}) at the last time step~$k$. 

\section{Simulation parameters}\label{appendix:parameters}
Here we give the details of the parameter values used in the simulations. To generate the locations $z_i[k]$ at every time $k$, we employ a simple movement model: For every individual~$i$, both entries of the initial $2\times 1$ location vector $z_i[1]$ are set to a uniform random number in $[0,1]$. Given the location vector $z_i[k]$ at any time $k$, we obtain the location vector at the next time $k+1$ as follows. With a probability of $p_\text{move}=0.1$, both entries of the location vector $z_i[k+1]$ are set to a uniform random number in~$[0,1]$. Otherwise, with a probability of $1-p_\text{move}= 0.9$, the location does not change, and hence $z_i[k+1]=z_i[k]$. To obtain the neighbourhoods $\mathcal{N}_i[k]$ from (\ref{N_all_from_d_inf}), we set the distance to $d_\text{inf} = 0.007$. A crucial metric for the qualitative epidemic behaviour is the epidemic threshold. If the effective infection rate $\tau=\beta/\delta$ is below the epidemic threshold, then the virus dies out rapidly and no individual is infectious any longer. Otherwise, above the epidemic threshold, a significant fraction of individuals is infected in the long run. For our simulations, we set the curing and infection probabilities to $\delta=0.25$ and $\beta=0.5$, respectively, which causes the effective infection rate $\tau$ to be above the epidemic threshold. Furthermore, we set the incubation probability to $\gamma=0.5$ and the fraction of asymptomatic infections to $\alpha=0.1$. The probability to contract a disease other than COVID-19 is set to $\vartheta=0.05$. For any individual $i$, the initial viral state is set to $X_i[1]=\mathcal{I}$ or $X_i[1]=\mathcal{I}_\text{a}$ with a probability of $0.01$ and $0.01\alpha$, respectively. Otherwise, with a probability of $(0.99-0.01\alpha)$, the initial viral state is set to $X_i[1]=\mathcal{S}$. Then, the prior distribution of the viral state $X_i[1]$ is given by $\operatorname{Pr}\left[ X_i[1] = \mathcal{S} \right] = 0.99-0.01\alpha$, $\operatorname{Pr}\left[X_i[1] = \mathcal{I}\right] =0.01$ and $\operatorname{Pr}\left[ X_i[1] = \mathcal{I}_\text{a}\right] =0.01\alpha$.


\begin{thebibliography}{10}

\bibitem{ferretti2020quantifying}
L.~Ferretti, C.~Wymant, M.~Kendall, L.~Zhao, A.~Nurtay, L.~Abeler-D{\"o}rner,
  M.~Parker, D.~Bonsall, and C.~Fraser, ``Quantifying {SARS-CoV-2} transmission
  suggests epidemic control with digital contact tracing,'' \emph{Science},
  vol. 368, no. 6491, 2020.

\bibitem{oliver2020mobile}  
N.~Oliver, B.~Lepri, H.~Sterly, R.~Lambiotte, S.~Deletaille, M.~De~Nadai,
  E.~Letouz{\'e}, A.~A. Salah, R.~Benjamins, C.~Cattuto,V.~Colizza, N.~de~Cordes, S.~P.~Fraiberger, T.~Koebe, S.~Lehmann, J.~Murillo, A.~Pentland, P.~N~Pham, F.~Pivetta, J.~Saramäki, S.~V.~Scarpino, M.~Tizzoni, S.~Verhulst, and P.~Vinc, ``Mobile
  phone data for informing public health actions across the {COVID-19} pandemic
  life cycle,'' \emph{Science Advances}, 2020.

\bibitem{drew2020rapid}
D.~A. Drew, L.~H. Nguyen, C.~J. Steves, C.~Menni, M.~Freydin, T.~Varsavsky,
  C.~H. Sudre, M.~J. Cardoso, S.~Ourselin, J.~Wolf, T.~D.~Spector, A.~T.~Chan, and COPE~Consortium, ``Rapid implementation of mobile technology for real-time epidemiology of
  {COVID-19},'' \emph{Science}, 2020.

\bibitem{funk2010modelling}
S.~Funk, M.~Salath{\'e}, and V.~A. Jansen, ``Modelling the influence of human
  behaviour on the spread of infectious diseases: a review,'' \emph{Journal of
  the Royal Society Interface}, vol.~7, no.~50, pp. 1247--1256, 2010.

\bibitem{kiss2010impact}
I.~Z. Kiss, J.~Cassell, M.~Recker, and P.~L. Simon, ``The impact of information
  transmission on epidemic outbreaks,'' \emph{Mathematical Biosciences}, vol.
  225, no.~1, pp. 1--10, 2010.

\bibitem{sahneh2011epidemic}
F.~D. Sahneh and C.~Scoglio, ``Epidemic spread in human networks,'' \emph{Proc.
  IEEE Conf. Decision Control}, pp. 3008--3013, 2011.

\bibitem{sahneh2012existence}
F.~D. Sahneh, F.~N. Chowdhury, and C.~M. Scoglio, ``On the existence of a
  threshold for preventive behavioral responses to suppress epidemic
  spreading,'' \emph{Scientific Reports}, vol.~2, p. 632, 2012.

\bibitem{theodorakopoulos2012selfish}
G.~Theodorakopoulos, J.-Y. Le~Boudec, and J.~S. Baras, ``Selfish response to
  epidemic propagation,'' \emph{IEEE Transactions on Automatic Control},
  vol.~58, no.~2, pp. 363--376, 2012.

\bibitem{schumm2013impact}
P.~Schumm, W.~Schumm, and C.~Scoglio, ``Impact of preventive behavioral
  responses to epidemics in rural regions,'' \emph{Procedia Computer Science},
  vol.~18, pp. 631--640, 2013.

\bibitem{funk2009spread}
S.~Funk, E.~Gilad, C.~Watkins, and V.~A. Jansen, ``The spread of awareness and
  its impact on epidemic outbreaks,'' \emph{Proceedings of the National Academy
  of Sciences}, vol. 106, no.~16, pp. 6872--6877, 2009.

\bibitem{tizzoni2014use}
M.~Tizzoni, P.~Bajardi, A.~Decuyper, G.~K.~K. King, C.~M. Schneider,
  V.~Blondel, Z.~Smoreda, M.~C. Gonz{\'a}lez, and V.~Colizza, ``On the use of
  human mobility proxies for modeling epidemics,'' \emph{PLoS Computational
  Biology}, vol.~10, no.~7, p. e1003716, 2014.

\bibitem{bengtsson2015using}
L.~Bengtsson, J.~Gaudart, X.~Lu, S.~Moore, E.~Wetter, K.~Sallah, S.~Rebaudet,
  and R.~Piarroux, ``Using mobile phone data to predict the spatial spread of
  cholera,'' \emph{Scientific reports}, vol.~5, p. 8923, 2015.

\bibitem{finger2016mobile}
F.~Finger, T.~Genolet, L.~Mari, G.~C. de~Magny, N.~M. Manga, A.~Rinaldo, and
  E.~Bertuzzo, ``Mobile phone data highlights the role of mass gatherings in
  the spreading of cholera outbreaks,'' \emph{Proceedings of the National
  Academy of Sciences}, vol. 113, no.~23, pp. 6421--6426, 2016.

\bibitem{UK_NHS_test_and_trace}
``Guidance for contacts of people with confirmed coronavirus (COVID-19) infection who do not live with the person,''
  \url{https://www.gov.uk/government/publications/guidance-for-contacts-of-people-with-possible-or-confirmed-coronavirus-covid-19-infection-who-do-not-live-with-the-person/guidance-for-contacts-of-people-with-possible-or-confirmed-coronavirus-covid-19-infection-who-do-not-live-with-the-person},
  accessed: 2020-08-25.

\bibitem{trevethan2017sensitivity}
R.~Trevethan, ``Sensitivity, specificity, and predictive values: foundations, pliabilities, and pitfalls in research and practice,'' \emph{Frontiers in Public Health}, vol.~5, p. 307, 2017.

\bibitem{donnelly1993correlation}
P.~Donnelly, ``The correlation structure of epidemic models,'' \emph{Mathematical Biosciences}, vol.~117, pp. 49--75, 1993.

\bibitem{cator2014nodal}
E.~Cator and P.~Van~Mieghem, ``Nodal infection in Markovian susceptible-infected-susceptible and susceptible-infected-removed epidemics on networks are non-negatively correlated,'' \emph{Physical Review E}, vol.~89, p. 052802, 2014.

\bibitem{du2000combinatorial}
D.~Du, and F.~Hwang, \emph{Combinatorial group testing and its applications}. World Scientific, 2020.

\bibitem{shental2020efficient}
N.~Shental, S.~Levy, V.~Wuvshet, S.~Skorniakov, B.~Shalem, A.~Ottolenghi, Y.~Greenshpan, R.~Steinberg, A.~Edri, R.~Gillis, M.~Goldhirsh, K.~Moscovici, S.~Sachren, L.~M.~Friedman, L.~Nesher, Y.~Shemer-Avni, A.~Porgador, and T.~Hertz, ``Efficient high-throughput {SARS-CoV-2} testing to detect asymptomatic carriers,'' \emph{Science Advances}, p. eabc5961, 2020.

\bibitem{van2014performance}
P.~Van~Mieghem, \emph{Performance {A}nalysis of {C}omplex {N}etworks and
  {S}ystems}.\hskip 1em plus 0.5em minus 0.4em\relax Cambridge University
  Press, 2014.

\bibitem{hong2013computing}
Y.~Hong, ``On computing the distribution function for the {P}oisson binomial
  distribution,'' \emph{Computational Statistics \& Data Analysis}, vol.~59,
  pp. 41--51, 2013.

\end{thebibliography}
\end{document}